\documentclass[twocolumn,english]{revtex4}
\usepackage[T1]{fontenc}
\usepackage[utf8]{luainputenc}
\usepackage{amsmath}
\usepackage{amssymb}
\usepackage{graphicx}
\usepackage{esint}

\makeatletter
\@ifundefined{textcolor}{}
{%
 \definecolor{BLACK}{gray}{0}
 \definecolor{WHITE}{gray}{1}
 \definecolor{RED}{rgb}{1,0,0}
 \definecolor{GREEN}{rgb}{0,1,0}
 \definecolor{BLUE}{rgb}{0,0,1}
 \definecolor{CYAN}{cmyk}{1,0,0,0}
 \definecolor{MAGENTA}{cmyk}{0,1,0,0}
 \definecolor{YELLOW}{cmyk}{0,0,1,0}
}

\usepackage{babel}

\usepackage{babel}

\usepackage{babel}

\usepackage{babel}

\usepackage{babel}

\usepackage{babel}

\usepackage{babel}

\makeatother

\usepackage{babel}
\begin{document}

\title{How to experimentally detect a GGE? - Universal Spectroscopic Signatures
of the GGE in the Tonks gas}

\author{Garry Goldstein and Natan Andrei}

\address{Department of Physics, Rutgers University}

\address{Piscataway, New Jersey 08854}
\begin{abstract}
In this work we study the equilibrium properties of the 1-D Lieb-Liniger
model in the infinite repulsion, Tonks-Girardeau regime. It is known
that for many initial states in the long time limit the Lieb-Liniger
gas equilibrates to the GGE ensemble. We are able to find explicit
formulas for the density density correlation functions the Tonks gas
in equilibrium. In the case that the initial and hence the final state
has low entropy per particle we find that the correlation function
has a universal form, in particular depends only on a finite number
of parameters corresponding to a finite set of ``key'' momenta and
has a power law dependance on distance. This provides a great experimental
signature for the GGE which may be readily measured through spectroscopy.
These signatures are universal so they robust to imperfections in
initial state preparation. 
\end{abstract}
\maketitle

\section{\label{sec:Introduction}Introduction}

Understanding the long time dynamics of a non-equilibrium one dimensional
system is a difficult task. The initial state is not a eigenstate
of the effective Hamiltonian but a complex superposition of such states.
As such the final state of the system does not depend on some eigenstate
and a few excitations but on a coherent superposition of various states.
If one wants to understand the emergence of a steady state one needs
to track the evolution of this coherent sum of states. This is the
problem that confronts theorists who wish to understand perturbed
quantum gases \cite{key-5,key-6}, ultrafast phenomena in superconductors
\cite{key-7} and thermalization in integrable systems \cite{key-8}.

One of the most surprising recent experimental and theoretical results
is that at long times the perturbed Lieb-Liniger gas \cite{key-9}
retains memory of its initial state \cite{key-5,key-10} and does
not appear to relax to thermodynamic equilibrium. This is due to the
fact that the Lieb Liniger hamiltonian: 
\begin{equation}
H_{LL}=\intop_{-\infty}^{\infty}dx\left\{ \partial_{x}b^{\dagger}\left(x\right)\partial_{x}b\left(x\right)+c\left(b^{\dagger}\left(x\right)b\left(x\right)\right)^{2}\right\} ,\label{eq:lieb-lin-hamiltonian}
\end{equation}
has an infinite number of conserved charges $I_{i}$. Here $b^{\dagger}\left(x\right)$
is the bosonic creation operator at the point $x$ and $c$ is the
coupling constant which in this work we will take to infinity. These
conserved quantities in turn imply that there is a complete system
of eigenstates for the Lieb Liniger gas which may be parametrized
by sets of rapidities $\left\{ k_{i}\right\} $. To understand the
equilibration of this gas it was recently proposed that it is insufficient
to consider only thermal ensembles but it is also necessary to include
these nontrivial conserved quantities. It was shown that the gas relaxes
to a state given by the generalized Gibbs ensemble GGE with its density
matrix being given by 
\begin{equation}
\rho_{GGE}=\frac{1}{Z}\exp\left(-\sum\alpha_{i}I_{i}\right)\label{eq:GGE_density_matrix}
\end{equation}

Where the $I_{i}$ are the conserved quantities given by $I_{i}\left|\left\{ k\right\} \right\rangle =\sum k^{i}\left|\left\{ k\right\} \right\rangle $
and the $\alpha_{i}$ are the generalized inverse temperatures and
$Z$ is a normalization constant insuring $Tr\left[\rho_{GGE}\right]=1$.
It was shown that correlation functions of the Lieb-Liniger gas at
long times may be computed by taking their expectation value with
respect to the GGE density matrix, e.g. $\left\langle \Theta\left(t\rightarrow\infty\right)\right\rangle =Tr\left[\rho_{GGE}\Theta\right]$.
It was also later shown \cite{key-1} that the the GGE ensemble is
equivalent to a pure state $\rho_{GGE}\cong\left|\vec{k}_{0}\right\rangle \left\langle \vec{k}_{0}\right|$
for an appropriately chosen $\left|\vec{k}_{0}\right\rangle $. It
is of great interest to provide some experimentally accessible signatures
for the GGE state, since most experimental signatures atleast in the
cold gases context focus on measurement of correlation functions we
will focus on these; in particular we focus on the simplest non-trivial
correlation function $\left\langle \rho\left(x\right)\rho\left(0\right)\right\rangle $.
Colloquially speaking there are two approaches towards using spectroscopic
signatures: 1) to determine the initial state with as great a precision
as possible from its measurable observables 2) to determine which
class of state it belongs to and to provide robust signatures of this
type of state. In some sense the second goal is more fundamental,
since the first presupposes knowledge of the second to some extend.
Furthermore the answer to the second question is usually universal
so it is robust to experimental imperfections. As such it will be
the focus of this work.

\begin{figure}
\begin{centering}
\includegraphics[width=1\columnwidth]{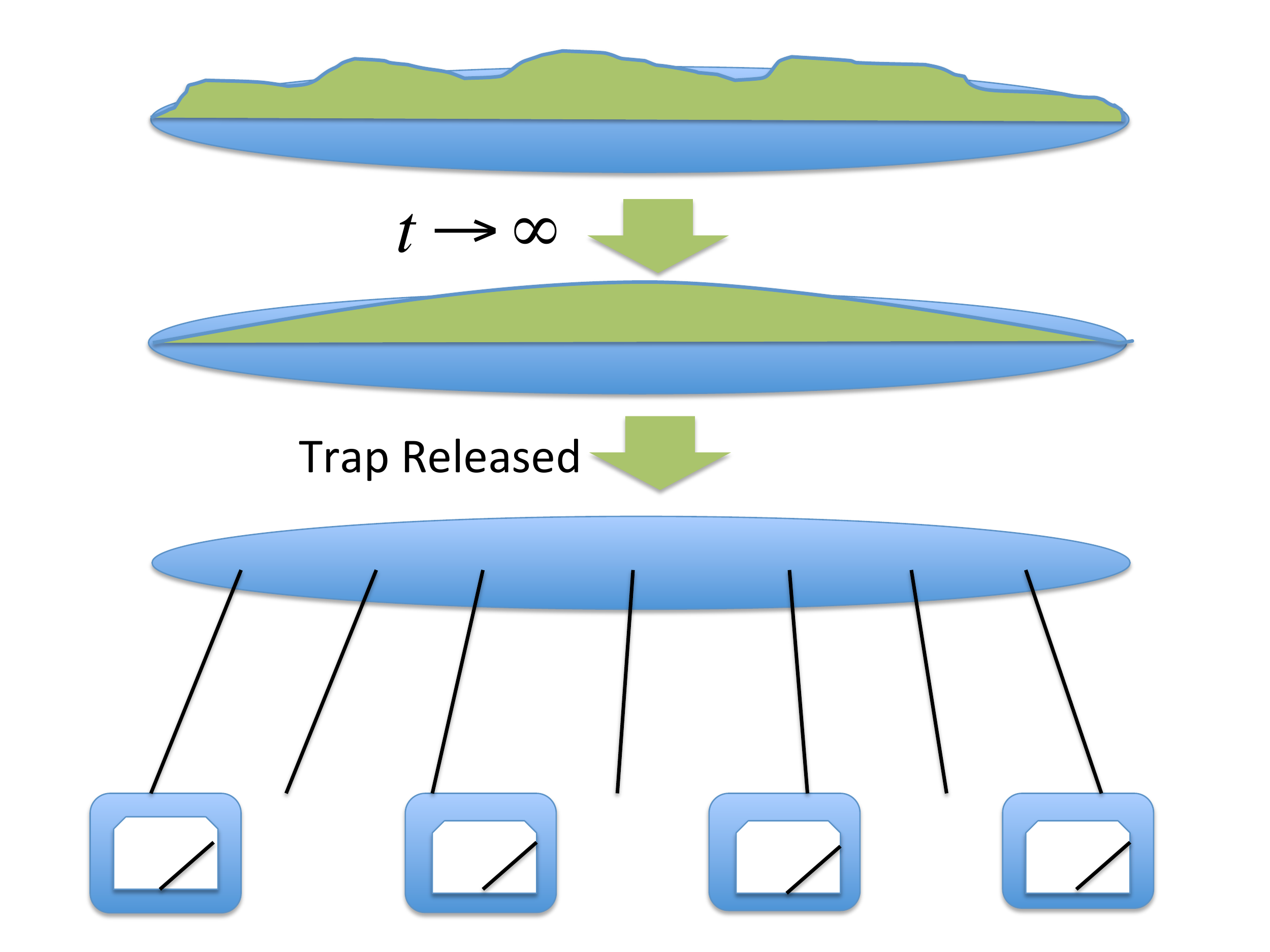}
\par\end{centering}

\protect\caption{\label{fig:Trap_release}The Lieb Liniger gas is initialized in some
nonequilibrium state, allowed to relax for a long time and its correlation
functions are measured.}
\end{figure}

\begin{figure}
\begin{centering}
\includegraphics[width=1\columnwidth]{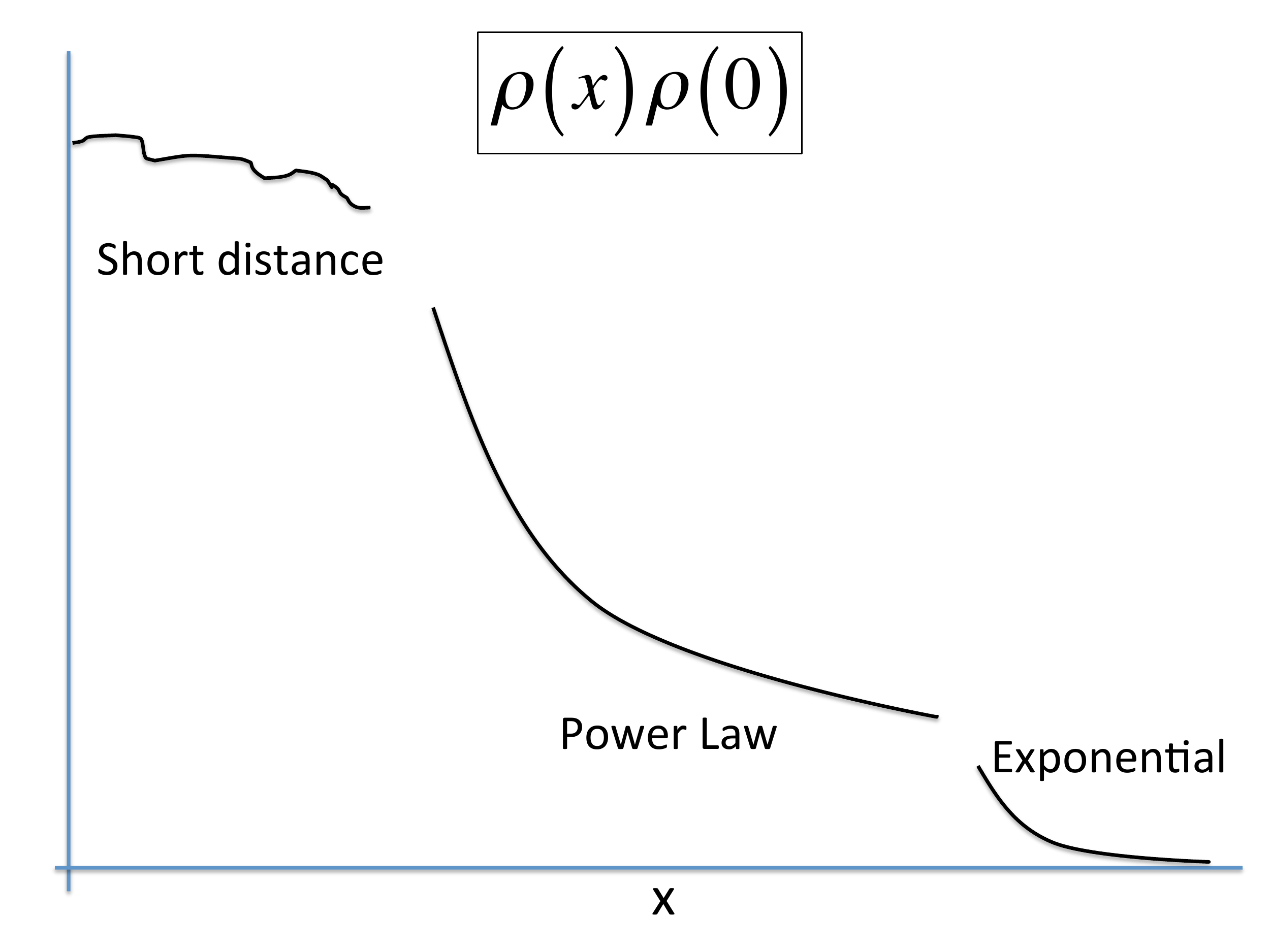}
\par\end{centering}

\protect\caption{\label{fig:Density_density}The density density correlation function
for the low entropy GGE. There is a complicated short distance region,
a power law decay $\sim\frac{1}{x^{2}}$ and an exponential decay
at long distances.}
\end{figure}

\begin{figure}
\begin{centering}
\includegraphics[width=1\columnwidth]{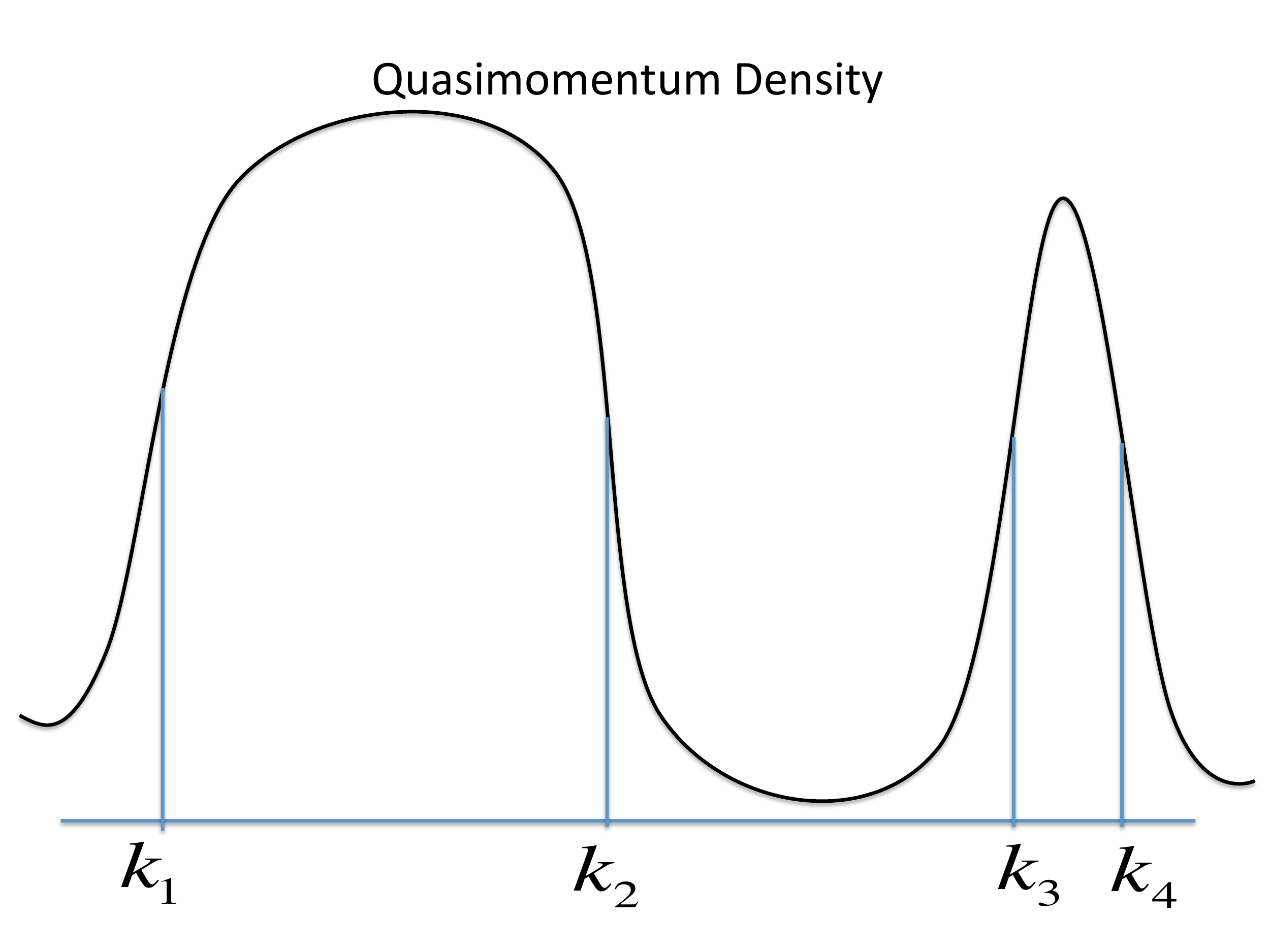}
\par\end{centering}

\protect\caption{\label{fig:Quansimomentud_density} The quasimomentum distribution
function $\rho_{p}\left(k\right)$. the points $k_{n}$ are determined
when $\rho_{p}\left(k_{n}\right)=\frac{1}{4\pi}$, for the low entropy
state these would correspond to rapid changes in the quasiparticle
density from $0\leftrightarrow\frac{1}{2\pi}$ see Section \ref{sec:Low-Entropy-per}.}
\end{figure}

In this work we consider the repulsive Lieb-Liniger model in the limit
of infinite interaction strength, Tonks Girardeau limit. We consider
the case when the gas was quenched from a non-equilibrium initial
state $\left|\Phi_{0}\right\rangle $ - which must be translationally
invariant and short ranged correlated - and allowed to relax for a
long time thereby establishing a GGE $\left|\Psi\right\rangle =e^{iH_{LL}t}\left|\Phi_{0}\right\rangle $
with $\left|\Psi\right\rangle \left\langle \Psi\right|\cong\rho_{GGE}$
see figure (\ref{fig:Trap_release}). In this regime when the GGE
is already established we compute the density density correlation
function of the gas and show that at long distances it is of universal
form - exponentially decaying $\sim e^{-\kappa x}$. This simple exponential
decay is a clear spectroscopic signature of the GGE. In the case when
the initial state, and hence the final state, has low entropy per
particle \cite{key-11} we find that at intermediate distances the
density density correlation function has a universal form it is a
power law decay $\sim\frac{1}{x^{2}}$, in particular the density
density correlation function has three ``regions'' and looks like
the one given in figure (\ref{fig:Density_density}). Furthermore
for any specific initial state with low entropy the exact density
density correlation function has a simple form depending only on the
the set of points $k_{n}$ such that the energy function vanishes
$\varepsilon_{\left\{ \alpha\right\} }\left(k\right)\equiv\sum\alpha_{i}k_{n}^{i}=0$.
The values $k_{n}$ are readily measurable experimentally through
time of flight imaging and are related to rapid changes in the particle
distribution function $\rho_{p}\left(k\right)$ discussed below, see
figure (\ref{fig:Quansimomentud_density}). Thus the density density
correlation functions give a finger print signature that the GGE has
been established. We also study a generalized case valid when the
initial state has long distance correlation functions \cite{key-10}
and the final sate of the gas is controlled by a generalized GGE.
We find that for the case when the initial state has low entropy per
particle that the long distance form of the density density correlation
function still has a power law form. However the prefactor is a complicated
$O\left(1\right)$ non-universal function. We would like to note that
the results pertaining to the exponential decay of correlations are
based on the assumption that the function $\varepsilon_{\left\{ \alpha\right\} }\left(k\right)$
is analytic in the complex plane which would happen for example when
the first few conserved quantities dominate the dynamics.

The rest of the paper is organized as follows: in Section \ref{sec:Reduction-to-a}
we simplify the GGE density matrix by reducing it to a single state
$\rho_{GGE}\cong\left|\vec{k}_{0}\right\rangle \left\langle \vec{k}_{0}\right|$;
in Section \ref{sec:Asymptotics-of-the} we present a general from
for the density density correlation function, we review the work done
in \cite{key-3}; in Section \ref{sec:Density-density-correlations}
we find the asymptotics of the density density correlation function
at long times and large distances for an arbitrary initial state,
we find an exponential decay of the correlation function; in Section
\ref{sec:Low-Entropy-per} we specialize to the case when there is
low entropy per particle and find a power law decay of the density
density correlation function at intermediate distances; in Section
\ref{sec:Bosonization} based on the results of Section \ref{sec:Low-Entropy-per}
we present a phenomenological Bosonization theory for the density
density correlations for the low entropy GGE state; in Section \ref{sec:Generalise-GGE}
we extend our results to the GGGE and in Section \ref{sec:Conclusions}
we conclude. 

\section{\label{sec:Reduction-to-a}Reduction to a pure state}

For the Lieb-Liniger gas it is possible to reduce the GGE density
matrix to a pure state. In particular it was shown that the GGE density
matrix corresponds to a specific pure state \cite{key-1}: that is
for any local correlation function $\Theta$:
\begin{equation}
tr\left[\Theta\rho_{GGE}\right]=\left\langle \vec{k}_{0}\right|\Theta\left|\vec{k}_{0}\right\rangle \label{eq:GGE_State_equivalence}
\end{equation}
for an appropriately chosen eigenstate of the Lieb Liniger Hamiltonian
$\left|\vec{k}_{0}\right\rangle $. Furthermore following the authors
of \cite{key-1} we are able to specify this pure state $\left|\vec{k}_{0}\right\rangle $.
To do so let us denote by $L\rho_{p}\left(k\right)dk$ as the number
of particles in the interval $\left[k,k+dk\right]$, $L\rho_{h}\left(k\right)dk$
as the number of holes in the interval $\left[k,k+dk\right]$ and
$L\rho_{t}\left(k\right)dk$ as the number of states in the interval
$\left[k,k+dk\right]$ so that $\rho_{t}\left(k\right)=\rho_{p}\left(k\right)+\rho_{h}\left(k\right)$.
Then the density of particles in k-space corresponding to $\left|\vec{k}_{0}\right\rangle $
is given by the following equations \cite{key-1}: 
\begin{equation}
\begin{array}{l}
2\pi\rho_{t}\left(k\right)=1+\int_{-\infty}^{\infty}K\left(k,q\right)\rho_{p}\left(k\right)\\
\varepsilon_{\left\{ \alpha\right\} }\left(k\right)-\varepsilon_{s}\left(k\right)-\frac{1}{2\pi}\int dqK\left(k,q\right)\ln\left(\varepsilon_{s}\left(q\right)\right)=0\\
\int\rho_{p}\left(k\right)dk=D
\end{array}\label{eq:Equation of state}
\end{equation}

Here $K\left(k,q\right)=\frac{2c}{c^{2}+\left(k-q\right)^{2}}$ and
we have introduced $\frac{\rho_{h}\left(k\right)}{\rho_{p}\left(k\right)}=e^{\varepsilon_{s}\left(k\right)}$,
$\varepsilon_{\left\{ \alpha\right\} }\left(k\right)=\sum\alpha_{i}k^{i}$
is the energy function of the GGE ensemble and we denote the density
of particles by $D$. In the case of the Tonks Girardeau gas this
relation between the GGE density matrix and the state $\left|\vec{k}_{0}\right\rangle $
greatly simplifies. In this case density corresponding to the state
$\left|k_{0}\right\rangle $ is given by 
\begin{equation}
\rho_{t}=\frac{1}{2\pi},\,\varepsilon_{s}\left(k\right)=\varepsilon_{\left\{ \alpha\right\} }\left(k\right)\equiv\varepsilon\left(k\right),\,\int\rho_{p}\left(k\right)dk=D\label{eq:Tonks-eqauation-state}
\end{equation}

From this it is possible to completely determine the density of particles
corresponding to the GGE density matrix, it is given by $\rho_{p}\left(k\right)=\frac{1}{2\pi}\frac{1}{1+e^{\varepsilon\left(k\right)}}$.
We note that the density function $\rho_{p}\left(k\right)$ is measurable
experimentally.

\section{\label{sec:Asymptotics-of-the}Asymptotics of the generating function}

As a preliminary step to understand the density density correlation
function we would like to study the generating function of density
correlations $\Theta_{\beta}\left(x\right)\equiv\exp\left(\beta\int_{0}^{x}b^{\dagger}\left(z\right)b\left(z\right)dz\right)$.
It is known \cite{key-2} that for a pure state $\left|\vec{k}_{0}\right\rangle $
the expectation value of $\Theta_{\beta}\left(x\right)$ is given
by the Fredholm determinant of the sine kernel with respect to the
state:
\begin{equation}
\begin{array}{l}
\left\langle \vec{k}_{0}\right|\Theta_{\beta}\left(x\right)\left|\vec{k}_{0}\right\rangle =\\
=\det\left(I+\frac{e^{\beta}-1}{\pi}\frac{1}{\sqrt{1+e^{\varepsilon\left(k\right)}}}\frac{\sin\left(k-q\right)\frac{x}{2}}{k-q}\frac{1}{\sqrt{1+e^{\varepsilon\left(q\right)}}}\right)
\end{array}\label{eq:Fredholm-determinant}
\end{equation}

The only dependance on $\left|\vec{k}_{0}\right\rangle $ comes from
$\varepsilon\left(k\right)\equiv\ln\left(\frac{\rho_{h}\left(k\right)}{\rho_{p}\left(k\right)}\right)$.
This Fredholm determinant has been well studied and the large distance
asymptotics of this determinant is given by \cite{key-3}:$ $
\begin{equation}
\left\langle \Theta_{\beta}\left(x\right)\right\rangle =\exp\left(\Delta\left(\nu\right)\right)\det\left(I-A\right)\label{eq:determinant_value}
\end{equation}

Here we have introduced the functions $\nu\left(k\right)=\frac{-1}{2\pi i}\ln\left(1+\frac{e^{\beta}-1}{1+e^{\varepsilon\left(k\right)}}\right)$,
$\Delta\left(\nu\right)=-\int_{-\infty}^{\infty}ix\nu\left(k\right)dk+\int_{-\infty}^{\infty}\int_{-\infty}^{\infty}\frac{\nu\left(k\right)\nu\left(q\right)}{\left(k-q+i0\right)^{2}}$
, $\alpha\left(k\right)=\exp\left(\int_{-\infty}^{\infty}\frac{\nu\left(q\right)}{q-k}dq\right)$
and $A=A^{-}A^{+}$ . Here $A_{jk}^{-}=\frac{h_{k}^{-}e^{-ixs_{k}^{-}}}{s_{j}^{+}-s_{k}^{-}}$,
$A_{jk}^{+}=\frac{h_{k}^{+}e^{ixs_{k}^{+}}}{s_{j}^{-}-s_{k}^{+}}$
with $h_{k}^{\pm}=\frac{\alpha\left(s_{k}^{\pm}\right)^{\mp2}}{\left(e^{\beta}-1\right)\left(\frac{1}{1+e^{\varepsilon\left(s_{k}^{\pm}\right)}}\right)^{2}e^{\varepsilon\left(s_{k}^{\pm}\right)}\varepsilon'\left(s_{k}^{\pm}\right)}$.
Here $s_{k}^{\pm}$ refer to the solutions of the equation $1+\frac{e^{\beta}-1}{1+e^{\varepsilon\left(s_{k}^{\pm}\right)}}=0$
with $\pm$ referring to solutions in the upper and lower half plane
respectively. This complex formula gives all the density density correlation
functions for the GGE. We will study and simplify this formula in
the remained of the paper. We note here that the number of solutions
$s_{k}^{\pm}$ might be infinite however because of the form of $A_{jk}^{\pm}$
only those near the real axis contribute to the correlation function
$\left\langle \Theta_{\beta}\left(x\right)\right\rangle $ so only
a finite number of terms matter. Corrections to this result, see Eq.
(\ref{eq:determinant_value}) decay faster then any exponential. We
note that this result depends on $\varepsilon\left(k\right)$ being
analytic in the complex plane, which would happen when the inverse
temperatures $\alpha_{i}$ converge fast enough.

\section{\label{sec:Density-density-correlations}Density density correlations}

We would like to specialize the above formulas, see Eq. (\ref{eq:determinant_value})
to the calculation of density density correlation functions. Since
$\left\langle \rho\left(x\right)\rho\left(0\right)\right\rangle =\frac{1}{2}\frac{d^{2}}{d\beta^{2}}\frac{d^{2}}{dx^{2}}\left\langle \Theta_{\beta}\left(x\right)\right\rangle _{\beta=0}$
it is sufficient to expand the above formulas for small $\beta$ upto
order $\beta^{2}$. With this in mind its sufficient to simplify $\nu\left(k\right)\cong i\frac{\beta}{2\pi}\frac{1}{1+e^{\varepsilon\left(k\right)}}$,
$\Delta\left(\nu\right)\cong-\beta x\int_{-\infty}^{\infty}\frac{1}{1+e^{\varepsilon\left(k\right)}}dk$,
and $\alpha\left(k\right)\cong1$. Furthermore introducing $\tilde{s}_{k}^{\pm}$
as solutions to the equations $\varepsilon\left(\tilde{s}_{k}^{\pm}\right)=\left(2n+1\right)\pi i$
closest to the points $s_{k}^{\pm}$ we have that $h_{k}^{\pm}\cong-\frac{\beta}{\varepsilon^{'}\left(\tilde{s}_{k}^{\pm}\right)}$
and therefore $ $$A_{jk}^{-}\cong-\frac{\beta}{\varepsilon^{'}\left(\tilde{s}_{k}^{-}\right)}\frac{e^{-ix\tilde{s}_{k}^{-}}}{\tilde{s}_{j}^{+}-\tilde{s}_{k}^{-}}$,
$A_{jk}^{+}\cong-\frac{\beta}{\varepsilon'\left(\tilde{s}_{k}^{+}\right)}\frac{e^{ix\tilde{s}_{k}^{+}}}{\tilde{s}_{j}^{-}-\tilde{s}_{k}^{+}}$.
We note that the points $\tilde{s}_{k}^{\pm}$ correspond to poles
of $\rho_{p}\left(k\right)=\frac{1}{1+e^{\varepsilon\left(k\right)}}$
which are close to solutions of the equation $1+\frac{e^{\beta}-1}{1+e^{\varepsilon\left(k\right)}}=0$
for small $\beta$. Now using $\det\left(I-A\right)\cong1-tr\left(A\right)$
and combining we get that 
\begin{equation}
\left\langle \rho\left(x\right)\rho\left(0\right)\right\rangle =\rho^{2}+\sum_{j,k}\frac{e^{ix\left(\tilde{s}_{j}^{+}-\tilde{s}_{k}^{-}\right)}}{\varepsilon'\left(\tilde{s}_{j}^{+}\right)\varepsilon^{'}\left(\tilde{s}_{k}^{-}\right)}=\rho^{2}+\left|\sum\frac{e^{ix\tilde{s}_{j}^{+}}}{\varepsilon'\left(\tilde{s}_{j}^{+}\right)}\right|^{2}\label{eq:simplified_densitydensity}
\end{equation}

The last equality comes about because of every pole of $\rho_{p}\left(k^{\star}\right)=\rho_{p}\left(k\right)^{\star}$
so a pole at $k$ corresponds to a pole at $k^{\star}$. For very
long distances only one of the exponentials dominates and we get that
$ $
\begin{equation}
\left\langle \rho\left(x\right)\rho\left(0\right)\right\rangle \cong\rho^{2}+\frac{e^{-2xIm\tilde{s}_{j}^{+}}}{\left|\varepsilon^{'}\left(\tilde{s}_{j}^{+}\right)\right|^{2}}\label{eq:Long-disnatnce_density density}
\end{equation}

As such at long distances the density density correlation function
for the Tonks gas after it has reached equilibrium generically has
a single exponential decay this result is robust to experimental imperfections.
This result compares well with the quench between free bosons and
hard core bosons studied in \cite{key-13}. We would like to note
that there are some cases where due to symmetry there are exactly
two degenerate minimal $\tilde{s}_{j}^{+}$. This happens for example
when $\varepsilon\left(k\right)=\varepsilon\left(-k\right)$. In this
case there are two minimal solutions $s_{1}$ and $-s_{1}^{\star}$.
Assuming that these are two distinct points, e.g. $s_{1}$ is not
purely imaginary we obtain that: 
\begin{equation}
\begin{array}{l}
\left\langle \rho\left(x\right)\rho\left(0\right)\right\rangle \cong\\
\cong\rho^{2}+4\frac{e^{-2xIms_{1}}}{\left|\varepsilon'\left(s_{1}\right)\right|^{2}}\sin^{2}\left(xRes_{1}-\ln\left(\frac{\varepsilon'\left(s_{1}\right)}{\left|\varepsilon'\left(s_{1}\right)\right|}\right)\right)
\end{array}\label{eq:simplified_density_density_two_minima}
\end{equation}
In this case, for a symmetric $\varepsilon\left(k\right)$, we see
that there is still an exponential decay but it is now multiplied
by $\sin^{2}\left(kx\right)$ with $k=Res_{1}$. This is an excellent
spectroscopic tool to determine the GGE as well as to test whether
the density of excitations $\rho_{p}\left(k\right)$ is symmetric
$\rho_{p}\left(k\right)=\rho_{p}\left(-k\right)$ or not. We note
though that the coefficient $Im\tilde{s}_{j}^{+}$ is very hard to
determine directly from a measurement of $\rho_{p}\left(k\right)$
because it requires an analytic continuation of the function $\rho_{p}\left(k\right)$
into the upper complex plane, it is measurable from interferometry
density density correlation measurement though. Below we shall see
that in the case that the initial state has low entropy per particle
for intermediate distance scales there is a different universal power
law behavior for the correlation functions.

\section{\label{sec:Low-Entropy-per}Low Entropy per Particle}

\begin{figure}
\begin{centering}
\includegraphics[width=1\columnwidth]{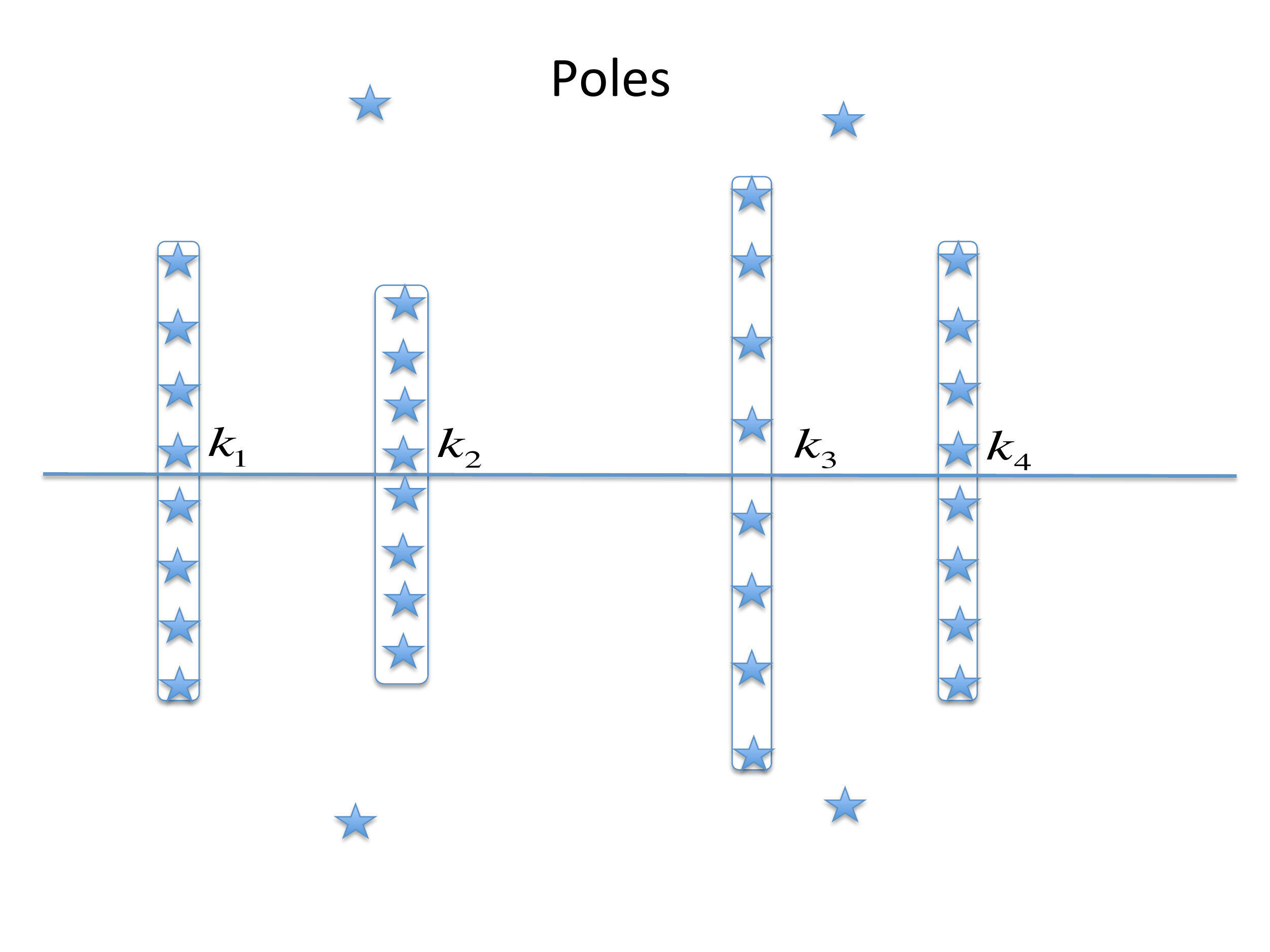}
\par\end{centering}

\protect\caption{\label{fig:Poles} Poles of the particle density function $\rho_{p}\left(k\right)$
in the complex plane. The poles relevant to the low entropy regime
near the points $k_{n}$ are circled.}
\end{figure}

We would like to consider the case where the final state has little
entropy per particle, or equivalently its at a low ``temperature''
\cite{key-12}. We know that the entropy per unit length is given
by $\frac{1}{2\pi}\int dk\left(\frac{1}{1+e^{\varepsilon\left(k\right)}}\ln\left(1+e^{\varepsilon\left(k\right)}\right)+\frac{1}{1+e^{-\varepsilon\left(k\right)}}\ln\left(1+e^{-\varepsilon\left(k\right)}\right)\right)$.
We see that this can only be small when either $\frac{1}{1+e^{\varepsilon\left(k\right)}}$
or $\frac{1}{1+e^{-\varepsilon\left(k\right)}}$ are small. From this
we see that $\left|\varepsilon\left(k\right)\right|\gg1$ for most
$\varepsilon\left(k\right)$ and that $\tilde{s}_{j}^{+}$ has a large
imaginary part except near specific values where $\varepsilon\left(k_{n}\right)=0$.
By considering the form of Eq. (\ref{eq:simplified_densitydensity})
we see that points with large imaginary values do not contribute so
we focus on the case where $\varepsilon\left(k_{n}\right)\cong0$.
Near such points $\varepsilon\left(k\right)\cong\left(k-k_{n}\right)\varepsilon'\left(k_{n}\right)$.
From this we get that the relevant solutions to $\varepsilon\left(\tilde{s}_{j}^{+}\right)=\left(2m+1\right)\pi i$
are given by $\tilde{s}_{n.m}^{+}\cong k_{n}+i\pi\frac{2m+1}{\varepsilon'\left(k_{n}\right)}$
with $m\geq0$ for $\varepsilon'\left(k_{n}\right)>0$ and $\tilde{s}_{n.m}^{+}\cong k_{n}-i\pi\frac{2m+1}{\varepsilon'\left(k_{n}\right)}$
with $m\geq0$ for $\varepsilon'\left(k_{n}\right)<0$ see figure
(\ref{fig:Poles}). Now considering Eq. (\ref{eq:simplified_densitydensity})
we see that the sum in the absolute value is given by: $\sum_{m}\frac{e^{ix\tilde{s}_{n.m}^{+}}}{\varepsilon'\left(\tilde{s}_{n,m}^{+}\right)}\cong\frac{e^{ik_{n}x}}{\varepsilon'\left(k_{n}\right)}\frac{e^{-\pi x/\left|\varepsilon'\left(k_{n}\right)\right|}}{1-e^{-2\pi x/\left|\varepsilon'\left(k_{n}\right)\right|}}\cong sgn\left(\varepsilon'\left(k_{n}\right)\right)\frac{e^{ik_{n}x}}{2\pi x}$.
From this we get that for intermediate distances in the low entropy
regime
\begin{equation}
\left\langle \rho\left(x\right)\rho\left(0\right)\right\rangle =\rho^{2}+\frac{1}{4\pi^{2}x^{2}}\left|\sum sgn\left(\varepsilon'\left(k_{n}\right)\right)e^{ik_{n}x}\right|^{2}\label{eq:intermediate distance}
\end{equation}
We note that this result is valid when there are many poles near the
points $k_{n}$ or alternatively $\left|\varepsilon'\left(k_{n}\right)\right|\gg1$.
We see that the density density correlation function in the Tonks
regime has universal form. It has a universal $\sim\frac{1}{x^{2}}$
dependance at intermediate distances with the exact prefactor depending
only on a few fixed momenta, which may be determined from the conserved
quantities $I_{i}$. We note that spectroscopically it is easier to
determine the momenta $k_{n}$ from the density function $\rho_{p}\left(k\right)$.
The points $k_{n}$ correspond to the points where $\rho_{p}\left(k_{n}\right)=\frac{1}{4\pi}$
which may be easily determined from time of flight imaging. Furthermore
these points are highly recognizable since at these points the density
function $\rho_{p}\left(k\right)$ has a rapid change $0\leftrightarrow\frac{1}{2\pi}$.
This result is universal and as such robust to experimental imperfections.
We note that for this section we do not need the analyticity of $\varepsilon\left(k\right)$
in the complex plane just near the real axis (which always happens).

\section{\label{sec:Bosonization}Bosonization}

Based on our results about the density density correlation function
for the low entropy Tonks gas see Eq. (\ref{eq:intermediate distance}),
we would like to Bosonize our gas. Previous schemes \cite{key-4}
to bosonize the Lieb Liniger gas were based on considering states
with low energy, those near the ground state. Here we would like to
bosonize low entropy but potentially high energy states. Following
\cite{key-4} we introduce $\phi_{l}\left(x\right)$ which counts
the number of particles up to the point $x$. Base on \cite{key-4}
we expect that
\begin{equation}
\rho\left(x\right)=\left[\rho_{0}-\frac{\nabla\phi\left(x\right)}{\pi}\right]\sum e^{ip\left(\rho_{0}-2\phi\left(x\right)\right)}\label{eq:generic_density}
\end{equation}
 Here the field $\phi_{l}\left(x\right)$ is rapidly varying and is
decomposed into $\phi_{l}\left(x\right)=2\pi\rho_{0}x-2\phi\left(x\right)$
with $\phi$ slowly varying. In our case for low entropy per particle
the density of bosons has rapid oscillations at the momenta $k_{n}$
with bosons being added when $\varepsilon'\left(k_{n}\right)<0$ and
bosons being removed when $\varepsilon'\left(k_{n}\right)>0$. The
simplest generalization of Eq. (\ref{eq:generic_density}) that includes
oscillations at the momenta $k_{n}$ which gives the correct leading
order contributions to the density and is translationally invariant
may be written as:$ $
\begin{equation}
\begin{array}{l}
\rho\left(x\right)=\rho_{0}-\sum_{k_{n}>k_{m}}\frac{1}{\pi\left(N-1\right)}\nabla\phi_{mn}\left(x\right)+\\
+A\sum_{k_{n}>k_{m}}sgn\left(\varepsilon'\left(k_{n}\right)\right)sgn\left(\varepsilon'\left(k_{m}\right)\right)\times\\
\qquad\times e^{i\left(k_{n}-k_{m}\right)x}e^{-2i\phi_{m,n}\left(x\right)}+h.c.
\end{array}\label{eq:Density_matrix}
\end{equation}
Where $\phi_{mn}$ are slowly varying field and $N$ is the total
number of solutions to the equation $\varepsilon\left(k_{n}\right)=0$.
We note that we have introduced a total of $\frac{N\left(N-1\right)}{2}$
bosonic fields each corresponding to an appropriate pair of points
$k_{n}$ and $k_{m}$ in the case of bosonizing the ground state this
would correspond to one field representing the left and right fermi
point. Furthermore the simplest action for the fields $\phi_{mn}$
may be written as \cite{key-4}:
\begin{equation}
S=\frac{1}{2\pi K}\sum_{k_{m}>k_{n}}\int dxd\tau\left(\frac{1}{u}\left(\partial_{\tau}\phi_{mn}\right)^{2}+u\left(\partial_{x}\phi_{mn}\right)^{2}\right)\label{eq:Action}
\end{equation}

This corresponds to a sum of independent actions for each bose field.
For simplicity we consider the zero temperature limit, it is the one
that reproduces the power law decay functions. This allows us to compute
the density density correlation function. It is given by:
\begin{equation}
\begin{array}{l}
\left\langle \rho\left(x,t\right)\rho\left(0,0\right)\right\rangle =\rho^{2}+\frac{NK}{4\pi^{2}}\frac{u^{2}t^{2}+x^{2}}{\left(x^{2}-u^{2}t^{2}\right)^{2}}+\\
\qquad+\frac{A}{\left(x^{2}-u^{2}t^{2}\right)^{K}}\left(\left|\sum sgn\left(\varepsilon'\left(k_{n}\right)\right)e^{ik_{n}x}\right|^{2}-N\right)
\end{array}\label{eq:Density_density_hypothesis}
\end{equation}

We hypothesize that this form with appropriate $u$ and $K$ is also
valid for the Lieb-Liniger gas for intermediate distances when considering
a GGE with relatively little entropy per particle. We would like to
note that this is only a conjecture and that there are many other
ways to bosonize the GGE Lieb liniger gas, for example introduce a
$K$ matrix and a $u$ matrix, however this is the simplest translationally
invariant way that reduces to known results about bosonizing the ground
state of the Lieb Liniger gas and given correct answers for the GGE.

\section{\label{sec:Generalise-GGE}Generalized GGE}

Recently it has been proposed that for states with long ranged correlation
functions the Lieb-Liniger gas does not equilibrate to the GGE but
to a generalized version of it, the GGGE \cite{key-10}. The density
matrix for the GGGE ensemble is given by $\rho_{GGGE}=\frac{1}{\tilde{Z}}\exp\left(-\sum\alpha_{i,j,k...}I_{i}I_{j}I_{k}...\right)$.
It was shown that for most purposes the GGGE density matrix is equivalent
to the diagonal ensemble density matrix \cite{key-10} that is $\rho_{GGGE}=\int d\vec{k}f\left(\vec{k}\right)\left|\vec{k}\right\rangle \left\langle \vec{k}\right|$,
for $f\left(\vec{k}\right)$ being a positive function of the momenta
$\vec{k}$ and $\int d\vec{k}f\left(\vec{k}\right)=1$. Here we would
like to show that in the case of initial states with low entropy per
particle the density density correlation function still retains its
$\sim\frac{1}{x^{2}}$ form. Indeed the density density function is
given by:
\begin{equation}
\begin{array}{l}
\left\langle \rho\left(x\right)\rho\left(0\right)\right\rangle =\int d\vec{k}f\left(\vec{k}\right)\left\langle \vec{k}\right|\rho\left(x\right)\rho\left(0\right)\left|\vec{k}\right\rangle =\\
=\rho^{2}+\frac{1}{x^{2}}\int d\vec{k}f\left(\vec{k}\right)\frac{1}{4\pi^{2}}\left|\sum sgn\left(\varepsilon'\left(k_{n}\right)\right)e^{ik_{n}x}\right|^{2}\\
\equiv\rho^{2}+\frac{C\left(x\right)}{x^{2}}
\end{array}\label{eq:Density_Density_GGGE}
\end{equation}

Because the integrand in the second line of Eq. (\ref{eq:Density_Density_GGGE})
is positive and has order one dependence on $x$ we notice that $C\left(x\right)=O\left(1\right)$.
We note that this equality strongly depends on the fact that $\left|\sum sgn\left(\varepsilon'\left(k_{n}\right)\right)e^{ik_{n}x}\right|^{2}\geq0$
otherwise averaging could lead to exponentially small results. Therefore
upto an order one non-universal order one function the density density
correlation function in the GGGE case has a $\sim\frac{1}{x^{2}}$
form for intermediate distances. This is a spectroscopic signature
of the GGGE. We note that in principle one could derive this result
starting from previous work \cite{key-10} however we have a much
more direct method.

\section{\label{sec:Conclusions}Conclusions}

We have studied the long time dynamics of a Tonks gas. Using the fact
that the gas equilibrates to the GGE we were able to find a pure state
corresponding to the Tonks gas. We computed exact density density
correlation functions for the equilibrated Tonks gas and found them
to have exponential decay at long distances. We considered the case
where the gas has little entropy per particle and found that at intermediate
distances the gas has a power law $\sim\frac{1}{x^{2}}$ density density
correlation function. We presented a bosonization hypothesis that
allowed us to extend our results to the Lieb Liniger gas with finite
interaction strength. In the future it would be interesting to confirm
this hypothesis as well as compute other correlation functions. We
note that our results provide clear spectroscopic signatures for identifying
a GGE using both the density function $\rho_{p}\left(k\right)$ and
the density density correlation function $\left\langle \rho\left(x\right)\rho\left(0\right)\right\rangle $.
In particular we show that in the low entropy per particle regime
the density density correlation has three ``regions'' which should
be easily measurable through interferometry. This result is universal
and as such is robust to experimental imperfections.

\textbf{Acknowledgments}: This research was supported by NSF grant
DMR 1006684 and Rutgers CMT fellowship.


\begin{thebibliography}{10}
\bibitem[11]{key-11} We note that entropy per particle is a dimensionless
quantity.
\bibitem[12]{key-12}We note that for the GGE ``temperature'' is
only a qualitative quantity.
\end{thebibliography}
\end{document}